\documentclass[10pt,letterpaper]{article}

\usepackage{iccv}
\usepackage{times}
\usepackage{epsfig}
\usepackage{graphicx}
\usepackage{amsmath}
\usepackage{amssymb}

\usepackage[pagebackref=true,breaklinks=true,letterpaper=true,colorlinks,bookmarks=false]{hyperref}

\iccvfinalcopy % *** Uncomment this line for the final submission

\def\iccvPaperID{****} % *** Enter the ICCV Paper ID here

\ificcvfinal\fi

\makeatletter
\DeclareRobustCommand\onedot{\futurelet\@let@token\@onedot}
\def\@onedot{\ifx\@let@token.\else.\null\fi\xspace}

\def\ie{\emph{i.e}\onedot}

\makeatother

\begin{document}

%%%%%%%%% TITLE
\title{Pandora3D: A Comprehensive Framework for High-Quality 3D Shape and Texture Generation} % PandoraX includes: Pandora3D, PandoraHuman, PandoraScene

% \author{Author, Author, Author, Author, Author, Author, Author, Author, Author\\
\author{Jiayu Yang$^1$\thanks{Equation contribution}, \quad 
    Taizhang Shang$^{1*}$, \quad
    Weixuan Sun$^{1*}$, \quad
    Xibin Song$^{1*}$, \quad
    Ziang Cheng$^{1*}$, \\
    Senbo Wang$^{1*}$, \quad
    Shenzhou Chen$^{1*}$, \quad
    Weizhe Liu$^{1*}$, \quad Hongdong Li$^{2}$, \quad Pan Ji$^{1}$\thanks{Project leader.}\\
$^{1}$ Tencent XR Vision Labs, $^{2}$ The Australian National University\\
% \and
}

\maketitle
\ificcvfinal\thispagestyle{empty}\fi

\begin{abstract}

This report presents a comprehensive framework for generating high-quality 3D shapes and textures from diverse input prompts, including single images, multi-view images, and text descriptions. The framework consists of 3D shape generation and texture generation. (1). The 3D shape generation pipeline employs a Variational Autoencoder (VAE) to encode implicit 3D geometries into a latent space and a diffusion network to generate latents conditioned on input prompts, with modifications to enhance model capacity. An alternative Artist-Created Mesh (AM) generation approach is also explored, yielding promising results for simpler geometries. (2). Texture generation involves a multi-stage process starting with frontal images generation followed by multi-view images generation, RGB-to-PBR texture conversion, and high-resolution multi-view texture refinement. A consistency scheduler is plugged into every stage, to enforce pixel-wise consistency among multi-view textures during inference, ensuring seamless integration. 

The pipeline demonstrates effective handling of diverse input formats, leveraging advanced neural architectures and novel methodologies to produce high-quality 3D content. This report details the system architecture, experimental results, and potential future directions to improve and expand the framework. The source code and pretrained weights are released at: \url{https://github.com/Tencent/Tencent-XR-3DGen}.

\end{abstract}
\section{Introduction}

Automated generation of high-quality digital 3D assets has drawn more and more attention in recent years. Digital 3D assets have become deeply ingrained in modern life and production. These assets vividly express the imaginations of creators across various fields, including gaming and film, bringing joy and creating immersive experiences for both players and audiences alike. Meanwhile, 3D assets also serve as essential building blocks in the domains of physical simulation and embodied AI, enabling machines and robots to understand the elements in the real world. However, the creation of 3D assets is far from simple; it is often a complex, time-consuming, and expensive process. Taking text prompts or an image as input, the digital 3D asset production pipeline commonly involves stages of 3D shape generation and texture generation, each requiring a high level of expertise and proficiency in digital content creation software.

In this report, we present Pandora3D, a framework designed for high-quality 3D shape and texture generation. The framework consists of two main components: 3D shape generation and texture generation.

\begin{itemize}
    \item 3D Shape Generation: The 3D shape generation pipeline utilizes a Variational Autoencoder (VAE) to encode implicit 3D geometries into a latent space. A diffusion network is then used to generate latents conditioned on input prompts, with modifications aimed at enhancing the model's capacity. We also explore an alternative Artist-Created Mesh (AM) generation approach, which shows promising results for simpler geometries.

    \item Texture Generation: The texture generation process is multi-staged, starting with the generation of frontal images, followed by multi-view images generation, RGB-to-PBR texture conversion, and high-resolution multi-view texture refinement. A novel consistency scheduler is integrated into every stage of this process to ensure pixel-wise consistency among multi-view textures during inference, leading to seamless integration.
\end{itemize}

The pipeline demonstrates effective handling of diverse input formats, leverages advanced neural architectures, and incorporates novel methodologies to produce high-quality 3D content. This report details the system architecture, experimental results, and potential future directions to improve and expand the framework.
\section{3D Shape Generation}
\label{sec:3dshape}

\subsection{3D Latent Space Diffusion} % weixuan, taizhang
The process begins by generating a 3D shape from a single image, multiple images, or a text prompt. This involves the following steps:
\begin{itemize}
    \item Variational Autoencoder (VAE): Compresses 3D geometries into a latent space, enabling efficient representation and processing.
    \item Diffusion Network: Generates latent representations conditioned on the input prompts. This network is adapted from CLAY~\cite{zhang2024clay} / Craftsman~\cite{li2024craftsman} / LAM3D~\cite{cui2024lam3d}, with modifications to improve the capacity and performance of the model. % taizhang
\end{itemize}

\subsubsection{Efficient 3D Geometry Autoencoder}
For 3D geometry compression model, we build upon CraftsMan~\cite{li2024craftsman}, which adopts structures introduced in  
3DShape2VecSet~\cite{zhang20233dshape2vecset} and Michelangelo~\cite{zhao2024michelangelo}. 
Furthermore, we leverage the multi-resolution training strategy proposed in CLAY~\cite{zhang2024clay}. 
This approach encodes 3D geometry into latent space by progressively sampling additional points from a 3D point cloud, which incrementally extends and refines the latent representation of the shape. Progressive sampling allows the model to focus on areas of higher geometric complexity, capturing both global structure and intricate details.
The primary goal of our VAE is to generate expressive latent embeddings that effectively guide the diffusion process in subsequent stages. To enhance the efficiency of this process, we propose a more advanced point sampling strategy. This method is designed to maximize the utility of the 3D point-cloud data by prioritizing points that contribute the most to capturing fine-grained features and spatial relationships.
This enhancement not only increases the model's capacity to handle large-scale data for improved scalability but also preserves the fine-grained details of 3D geometry.

\begin{figure*}
    \centering 
    \includegraphics[width=0.9\linewidth]{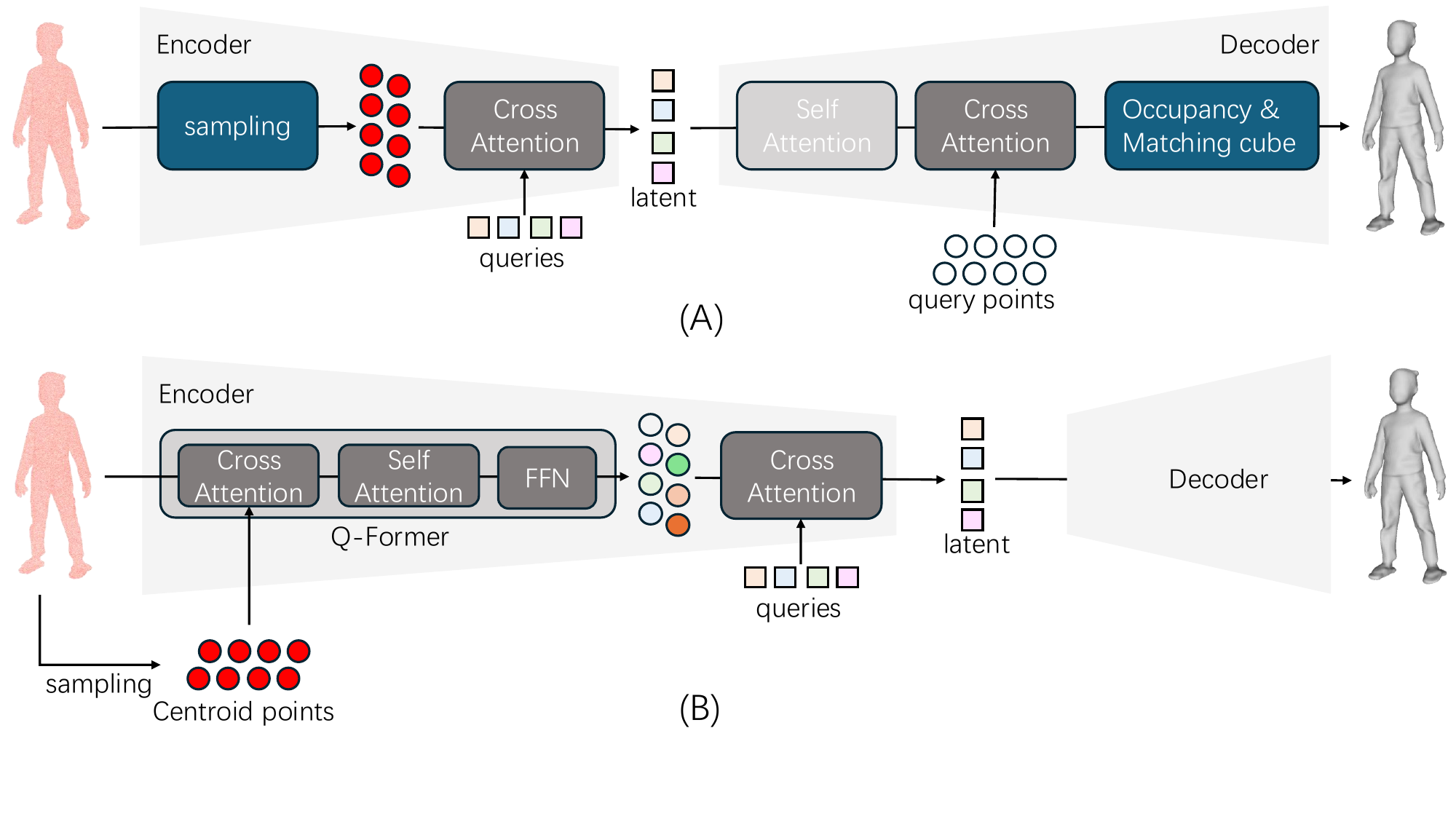}
    \caption[3D Geometry variational autoencoder.]
    {3D Geometry variational autoencoder. (A): Our base VAE for 3D geometry compression. (B) Extended VAE for efficient 3D geometry compression.}
    \label{fig:vae}
\end{figure*}

The design options of our VAE are illustrated in Fig.~~\ref{fig:vae}. We employ the model structure introduced in 3DShape2VecSet as our base model. This approach involves embedding the input point cloud, augmented with normal information $X \in \mathbb{R}^{N \times 6} $, which is sampled from a mesh $M$, into a latent code using a learnable embedding function and a cross-attention encoding module:

\begin{equation}
Z = \varepsilon(X) = CrossAttn(q, PosEmb(X)) \;,
\end{equation}
where $q \in \mathbb{R}^{m \times d}$ represents a set of learnable queries that compress the sampled points into a latent embedding. The cross-attention mechanism ensures effective integration of geometric and positional features into the latent space. The VAE’s decoder is composed of successive self-attention layers followed by a cross-attention layer. The cross-attention layer maps the latent embeddings back into 3D geometry, enabling reconstruction:

\begin{equation}
D(Z, p) = CrossAttn(PosEmb(p), SelfAttn(Z)) \;,\label{eq:decoder}
\end{equation}
where $p$ denotes random query points in 3D space, these points query with the latent and output occupancy logits.
This base VAE implementation is illustrated in Fig.~\ref{fig:vae} (A).

Following the approach outlined in CLAY~\cite{zhang2024clay}, we adopt a multi-resolution training strategy to progressively upscale the model's capacity. Specifically, we incrementally increase the number of sampled points from 4096 to 32768 while simultaneously extending the latent embedding dimensionality from 256 to 2048. This progressive training scheme gradually introduces more detailed input information to the model, enabling it to capture finer geometric details. At the same time, the expanded latent embedding length increases the model’s capacity to represent complex features. Together, these enhancements enrich the latent space, thereby providing a more robust foundation for the subsequent diffusion model training. This multi-resolution approach ensures an efficient and scalable training process, optimizing both the model's performance and its ability to generalize across diverse 3D geometries.

% This multi-resolution approach ensures an efficient and scalable training process, optimizing both the model's performance and its ability to generalize across diverse 3D geometries.

% Such a training scheme gradationally increases the input information amount and analogously increases capacity of the latent embeddings, which can enrich the following diffusion model training.

Recall that the primary objective of our VAE is to generate expressive latent embeddings. While the previously mentioned approach progressively increases the number of sampling points, each object contains a total of 500k points, leaving many points unsampled. This results in inevitable information loss, as not all geometric details are captured in the latent representation.
Furthest point sampling~\cite{qi2017pointnet++} has the potential to mitigate this issue by selecting more representative points. However, this method is significantly slower compared to random sampling, making it less practical for large-scale training scenarios. This residual information loss can pose challenges during the diffusion process, as it may hinder the generation of high-quality latent embeddings. Consequently, the decoder is tasked with reconstructing fine-grained 3D details that might not be adequately represented in the latent embedding, potentially limiting the overall quality and fidelity of the reconstructed geometry.

% Recall that our VAE's objective is to produce expressive latent embeddings. 
% Although the aforementioned method gradationally increases sampling points amount, for each object we have in total 500k points so 
% there are still unsampled points that lead to inevitably information loss.
% Furthest point sampling~\cite{qi2017pointnet++} might remit the problem but it is significantly slower than random sampling.
% Such information loss may hinder the diffusion process from generating high-quality latent and require the decoder to reconstruct the fine-grained 3D details that might not exist in the latent embedding.

We have developed an enhancement to our Variational Autoencoder (VAE) model that allows it to operate without sampling, while still retaining all data points. A straightforward approach might involve utilizing PointNet++ \cite{qi2017pointnet++} to compress features from a point cloud into a few "centroid points" through the use of cascaded convolutional layers. However, this method demands a substantial amount of memory, especially when managing point cloud data consisting of millions of points. To address this, our model optimizes the processing of large-scale point clouds more efficiently, reducing the memory burden without compromising the integrity and richness of the data.
Alternatively, we opt to sample a set of centroid points  $M \in \mathbb{R}^{m \times 6}$ and employ a Q-former~\cite{li2022blip} style module to compress the raw point cloud data onto these centroids. The core component of the Q-former, the cross-attention mechanism, exhibits a computational complexity $O(2MNd)$, where $M$ is the amount of centroids and $N$ is the size of the input point cloud and $d$ is feature dimension.
Although utilizing a memory-efficient attention method such as FlashAttention~\cite{dao2022flashattention} helps, it remains resource-intensive and slow for processing large point clouds directly without sampling. To overcome these challenges, we propose the adoption of a linear attention mechanism~\cite{qin2024various,qin2023transnormerllm} for implementing cross-attention within our Q-former module.
The theoretical complexity of this approach is $O(Md^2+Nd^2)$. 
Given that $N>>N>>d$, the computational load of linear cross-attention is significantly reduced compared to traditional cross-attention methods. During the training of our VAE, we randomly select points from the original point clouds to serve as centroid points. These centroids, which have a dimension larger than 6, act as queries in the Q-former and compress geometric information from the raw point cloud. Subsequently, these centroid points are processed by the VAE encoder to generate latent embeddings.
This extended VAE is depicted in Fig.~\ref{fig:vae} (B). 

Similarly, we still adopt multi-reolustion training strategy to progressively increase the centroid points amount and latent embedding length to enlarge the latent embedding capacity. 
In addition, we empirically find that progressively increasing the training data amount can accelerate model convergence.
Methods like 3DShape2VecSet and CLAY, which derive latents from sampled points of the original point cloud, inevitably suffer from detail loss. 
Our extension effectively addresses this issue of information loss and maximizes the utilization of high-resolution point clouds, thereby preserving more detailed and accurate representations.

% We randomly sample
% In addition, inspired by 3Dshape2vecset~\cite{zhang20233dshape2vecset}, we want a set of proper centroid points that can 

\subsubsection{Diffusion Pipeline}

\begin{figure*}
    \centering 
    \includegraphics[width=0.9\linewidth]{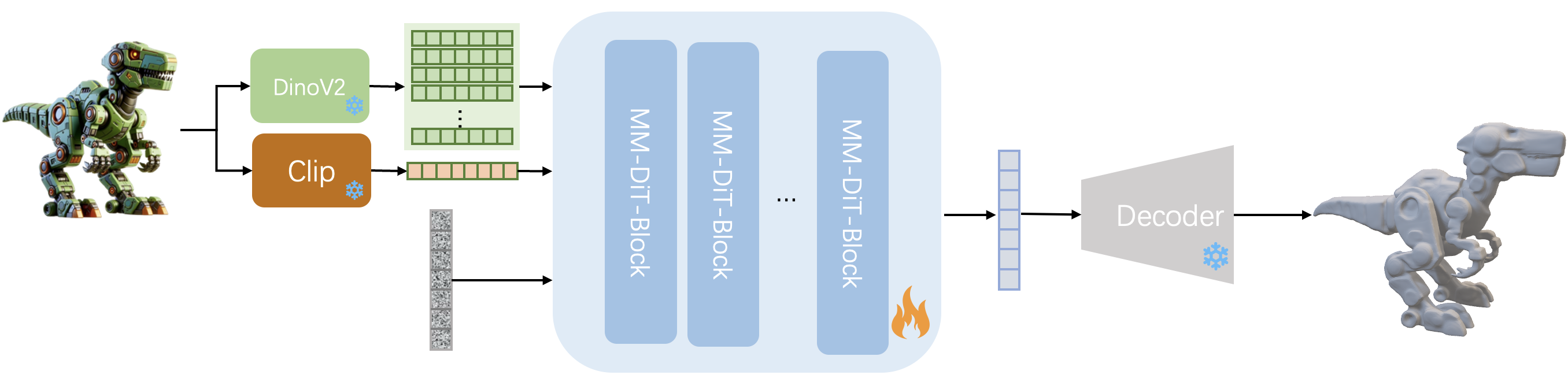}
    \caption[Diffusion pipline.]
    {Diffusion pipline. In the process of training a diffusion model, the DinoV2, CLIP, and VAE Decoder components are kept frozen}
    \label{fig:diffusion:1}
\end{figure*}

The diffusion pipeline is illustrated in Fig.~\ref{fig:diffusion:1}. We employ Multimodel Diffusion Transformer(MMDiT)~\cite{esser2024scaling} as our diffusion backbone, utilizing two pretrained models, specifically, CLIP-ViT-L/14~\cite{radford2021learning} as the global image feature extractor, and Dino-V2-Large~\cite{oquab2023dinov2} for local image feature extraction. Instead of employing DDPM, we utilize the flow matching schedule. Following CLAY~\cite{zhang2024clay} methodology, the diffusion model is trained in coarse-to-fine manner.

To enhance the image control effect, we use both global and local image feature as condition features of diffusion model. Global condition feature $z_{global} \in \mathbb{R}^L$ is extracted with ClIP vision encoder, meanwhile, local detail condition feature $z_{local} \in \mathbb{R}^{L \times 1024}$ is extracted with Dino vision encoder. The global condition and local condition are integrated into the diffusion model through MMDiT Block following Stable Diffusion 3~\cite{esser2024scaling}. The diffusion model we use has 2.3B parameters and 28 layers MMDiT block.

To enhance practical utility in 3D design workflows, we have extended our geometry generation framework to accept multi-view conditional inputs. This architectural advancement enables finer-grained geometric control through multi-view visual guidance. The system accommodates variable numbers of reference images (stochastically sampled from 1 to 4 views per instance) within a unified architecture, eliminating requirement for fixed-size input configurations. All synthesized geometries maintain spatial alignment with the primary view's coordinate system (defined by the first input image). Input images must be arranged in ascending azimuth order $\{\theta_1, \theta_2, \ldots, \theta_n\} \quad$ where $\quad \theta_i \in [0^\circ, 360^\circ)$. Multiview feature representations are aggregated through ordered concatenation along the sequence dimension, preserving relative spatial-semantic correspondence across views. Accelerated convergence is achieved via progressive transfer learning, where parameters initialized from our single-view conditioned model undergo fine-tuning using multiview datasets while maintaining pretrained backbone weights during initial phases.

\subsection{Meshing and UV Unwrapping} % zacheng
Once the 3D geometry is generated, it undergoes isosurface extraction, remeshing and UV unwrapping so that a texture-ready triangle mesh is produced.

\subsubsection{Isosurface extraction}
We perform a modified version of marching cubes algorithm~\cite{lorensen1998marching,nielson2003marching} to efficiently extract a watertight mesh from geometry tokens.

Marching cubes traditionally require a dense occupancy grid of $D\times D \times D$ occupancy values. Directly computing such a dense grid with Eq.~\eqref{eq:decoder} incurs a $O(D^3)$ time complexity that is prohibitively expensive at high resolutions $D$. To improve efficiency, we adopt a coarse to fine strategy: starting from a coarse grid resolution $d_0\ll D$, we iteratively build a sparse finer grid of resolution $d_{i+1}=2 d_i$ whose cells are subdivided from active cells in the coarser grid of resolution $d_i$ that are close to the isosurface. This strategy ensures that most occupancy queries of Eq.~\eqref{eq:decoder} are confined within a small margin around the isosurface and significantly reduces the number of queries required for isosurface extraction, achieving two to three orders faster mesh extraction. 

To guarantee a watertight mesh, at the highest level $d_n=D$, we expand the sparse active cells along the isosurface to eliminate holes and perform Lewiner's topology check~\cite{lewiner2003efficient} to ensure manifoldness. We implement the sparse marching cubes as a custom CUDA kernel function to maximize efficiency.

\subsection{Remesh and UV unwrap} 
The triangle meshes extracted from Marching cubes may contain poorly constructed elements such as collapsed faces or slivers. Furthermore, they often exhibit a high face count that could create problems for downstream applications. We overcome these issues with an optional remeshing step using either an off-the-shelf quad-remesher\footnote{\url{https://exoside.com/}} or isotropic remeshing~\cite{pietroni2009almost} followed by QEM triangle decimation~\cite{garland1997surface}. 

In addition, we use the open source project UV-Atlas~\cite{zhou2004iso} for UV charting and packing. At this point, we obtain a polygon mesh that is ready for texture generation.

\subsection{Alternative Approach: Artist-Created Meshes Generation} % weizhe
An alternative approach we explored involves directly generating the mesh, bypassing the initial generation of geometry as an implicit function followed by mesh extraction. This method effectively produces meshes with reasonable topology, akin to those crafted by artists for simple shapes. However, it encounters difficulties when applied to complex geometries, where maintaining structural integrity and topological accuracy becomes challenging.

\subsubsection{Mesh Compression}

Direct regression of vertex coordinates results in substantial memory consumption, which consequently limits the number of faces the model can handle. To mitigate this issue, we adopt the methodology proposed by BPT~\cite{weng2024scaling}, which involves compressing the original vertex coordinates using block index compression and patchified aggregation. Specifically, for a vertex $v_{i} = (x_i,y_i,z_i)$, the block-wise indexing $(b_i,o_i)$ is formulated as follows:
\begin{equation}\label{eq:bpt1}
\begin{split}
    b_{i} &= (x_i|O)\cdot B^{2} + (y_{i}| O)\cdot B + z_{i}|O\;, \\
    o_{i} &= (x_{i}\%O) \cdot O^{2} + (y_{i}\%O)\cdot O + z_{i}\%O\;.
\end{split}
\end{equation}
In this formulation, the symbols $|$ and $\%$ represent division without remainder and the modulo operation, respectively. This approach segments the coordinates along each axis into $B$ blocks, each of length $O$. To further enhance the compression ratio, we employ the patchified aggregation technique as described in~\cite{weng2024scaling}. This technique aggregates the faces connected to the same vertex into a non-overlapping patch and utilizes dual-block indices to denote the starting point of a patch. Consequently, the offset vocabulary is shared between the center vertex and the surrounding vertices. The center patch is formulated as follows:
\begin{equation}\label{eq:bpt2}
P_{c} = (b'_{c},o_{c},b_{1},o_{1},b_{2},o_{2},\dots,b_{n},o_{n})\;.
\end{equation}
In this context, $b'_{c}$ and $o_{c}$ denote the blocking index and the offset index of the center patch, respectively. These indices are critical for accurately referencing the spatial configuration of the patch within the compressed data structure.

To achieve this, we initially convert vertex coordinates into discrete values with a resolution of $R$. Subsequently, we encode the mesh information, including vertices and faces, into a discrete token sequence. This sequence can be decoded back into a mesh using the same technique. It is important to note that this encoding and decoding process is governed by predefined rules and does not involve any learnable parameters.

\subsubsection{Autoregressive Model for Mesh Generation}
\begin{figure*}
    \centering 
    \includegraphics[width=0.9\linewidth]{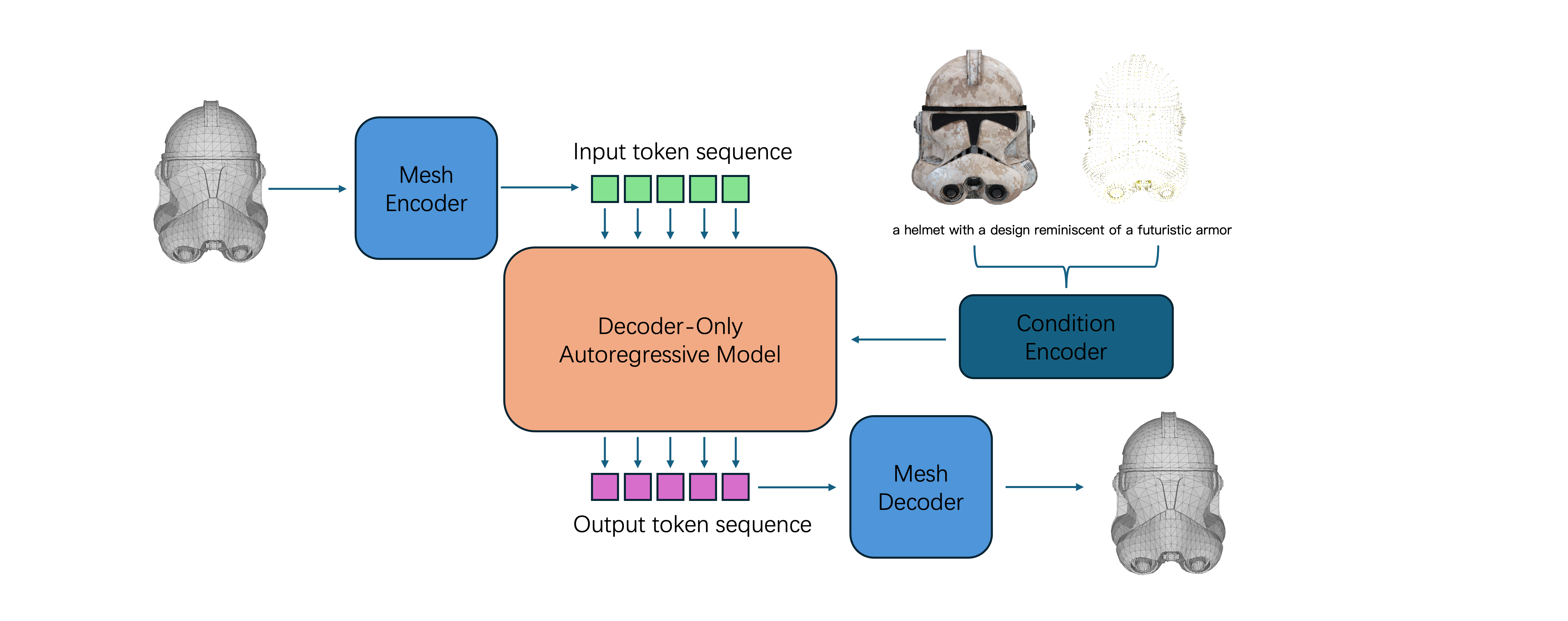}
    \caption[Pipeline for Artist-Created Meshes Generation.]
    {Pipeline for Artist-Created Mesh Generation. Initially, meshes are encoded into discrete token sequences. These sequences are then processed through a decoder-only autoregressive model that utilizes a Transformer network architecture. To enforce multi-modality condition control, a pretrained condition encoder network is employed. This network effectively integrates diverse modalities, ensuring that the generated meshes adhere to specified conditions.}
    \label{fig:am:1}
\end{figure*}

In this section, we describe the methodology for generating novel shapes from various modalities using the compression technique outlined in the preceding sections. Fig.~\ref{fig:am:1} illustrates the pipeline of our approach. Initially, a mesh is encoded into discrete token sequences utilizing the method detailed previously. Subsequently, a decoder-only autoregressive model is employed to predict subsequent tokens based on preceding ones. To facilitate multi-modality condition control, a pre-trained condition encoder network is utilized to encode condition information, such as images, text, and point clouds, into latent features. These features serve as the contextual input for the decoder-only model. The resulting token sequence can then be decoded back into the final mesh using a mesh decoder. It is important to note that both the mesh encoder and mesh decoder are purely rule-based, as previously explained, and do not involve any learnable parameters.

\begin{figure*}
    \centering 
    \includegraphics[width=0.9\linewidth]{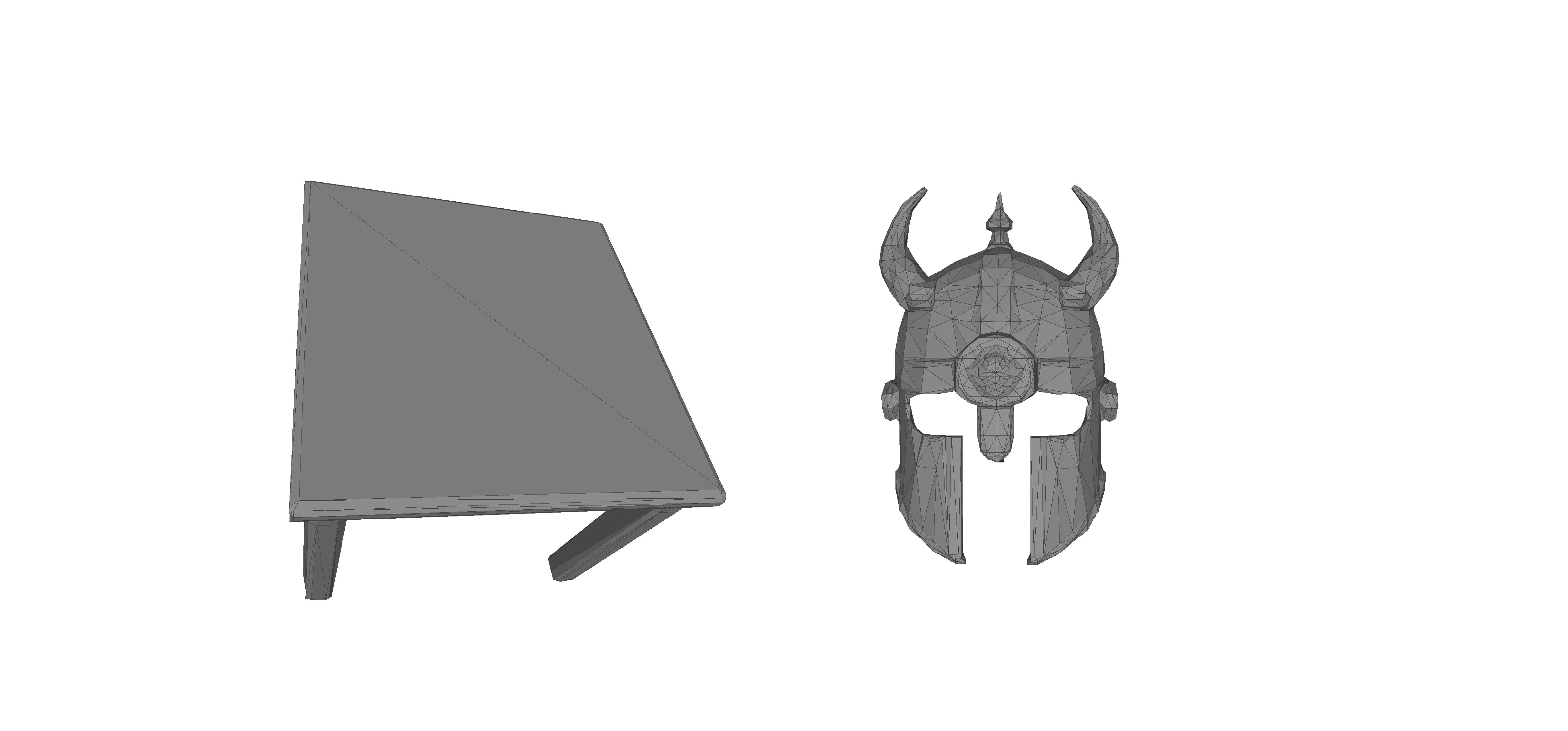}
    \caption[Example Meshes Generated by Our Artist-Created Meshes Generation Model.]
    {Example Meshes Generated by Our Artist-Created Mesh Generation Model. The meshes produced by our model demonstrate superior performance in maintaining topological consistency, showcasing the effectiveness of our approach in generating high-quality artistic meshes.}
    \label{fig:am:2}
\end{figure*}

Fig.~\ref{fig:am:2} visualizes the meshes generated by our model, which exhibit superior topology consistency with a minimal number of faces.

\section{Texture Generation}
\label{sec:texture}
\begin{figure*}
    \centering 
    \includegraphics[width=\linewidth]{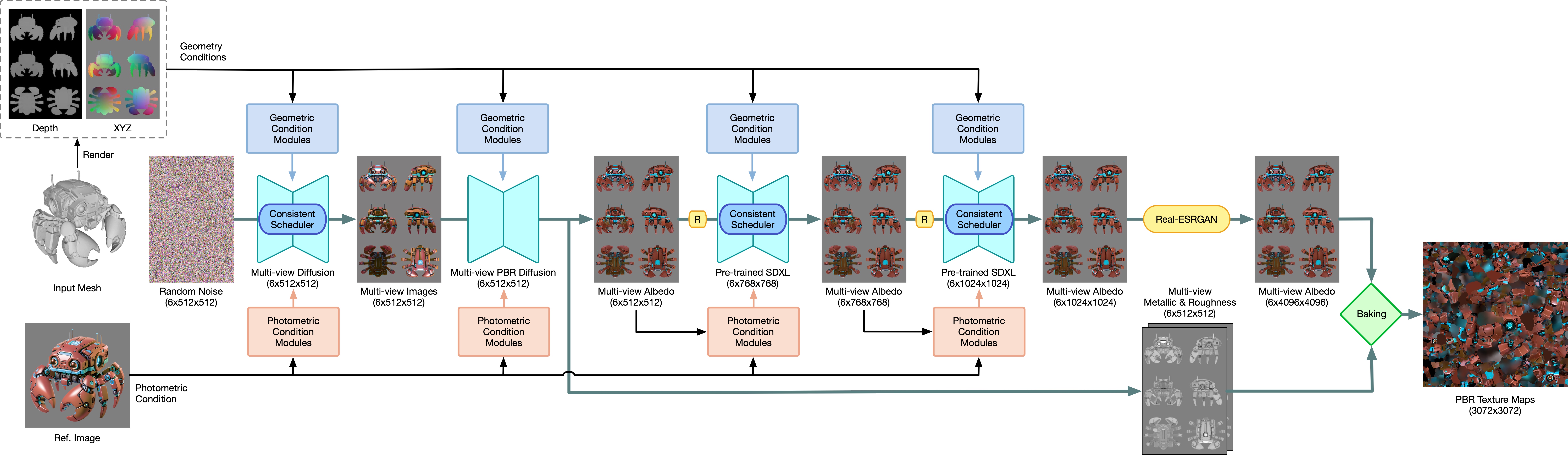}
    \caption[Texture Generation Pipeline]
    {Texture Generation Pipeline (input image and mesh from Trellis3D). }
    \label{fig:texture_pipeline}
\end{figure*}

The proposed texture generation pipeline consists of several stages, each contributing to the generation of consistent and high-quality textures. Fig.~\ref{fig:texture_pipeline} illustrates the texture generation pipeline. The pipeline begins with a 3D mesh without texture. Below we introduce each stage in detail.

\subsection{Frontal Image Generation} % shenzhou
\begin{figure*}
    \centering 
    \includegraphics[width=0.9\linewidth]{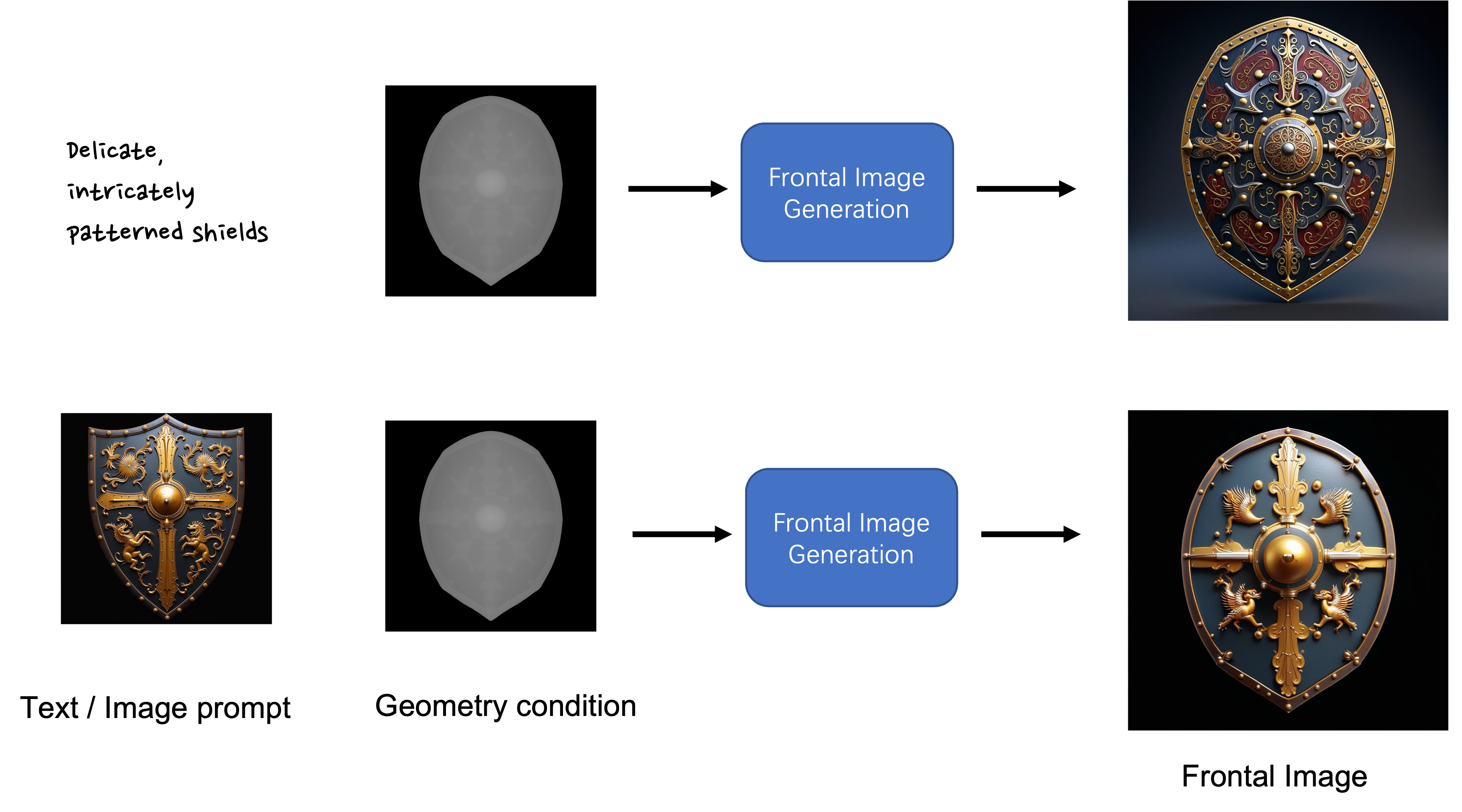}
    \caption[Frontal Image Generation.]
    {Both textual and visual prompts are transformed into a frontal image that is aligned with the frontal-view geometry.}
    \label{fig:frontal}
\end{figure*}

If the input prompt is text, a frontal image is initially generated conditioned on a depth map derived from the 3D geometry. This process involves rendering the 3D mesh into a depth map and utilizing depth-conditioned diffusion models \cite{zhang2023adding} to produce the frontal image. Alternatively, if the input is an image, we integrate the IP-Adapter \cite{ye2023ip} and ControlNet \cite{zhang2023adding} to generate the frontal image. As illustrated in Fig. \ref{fig:frontal}, both text and image prompts are converted into a geometry-aligned frontal image, which serves as the input for subsequent texture generation. 

\subsection{Multi-view RGB Image Generation} % xibin
\label{sec:texture:multiview}
A single-view to multi-view image generator creates multi-view RGB images conditioned on the multi-view position maps and the frontal image. Please note that normal and depth maps can also be used here. We first train a multi-view image generator with a network structure similar with Zero123++~\cite{shi2023zero123++}, then, we combine the ControlNet~\cite{zhang2023adding} with Zero123++~\cite{shi2023zero123++} conditioned on the position (XYZ coordinate) maps, enabling the generation of position-aligned multi-view images. Whether the frontal image originates from text and depth or is provided as input, the multi-view image generator generates six multi-view images (each $512 \times 512$) starting from random Gaussian noise, with geometric conditions and photometric modules. An example is shown in the first image in Fig. \ref{fig:multiviewpbr}.

% \begin{figure*}
%     \centering 
%     \includegraphics[width=0.3\linewidth]{figures/texture/final_images_d2rgb.png}
%     \caption[Multi-view RGB image]
%     {Multi-view RGB image.}
%     \label{fig:multiviewrgb}
% \end{figure*}

\subsection{Multi-view PBR Image Generation} % xibin
% The generated multi-view RGB images are decomposed into multi-view PBR (Physically Based Rendering) images. This stage includes generating multi-view albedo, metallic, and roughness maps (each 512x512), utilizing multi-view depth, XYZ coordinate map, frontal images and previously estimated multi-view RGB images as additional conditioning inputs through photometric and geometric condition modules.

\begin{figure*}
    \centering
    \includegraphics[width=0.24\linewidth]{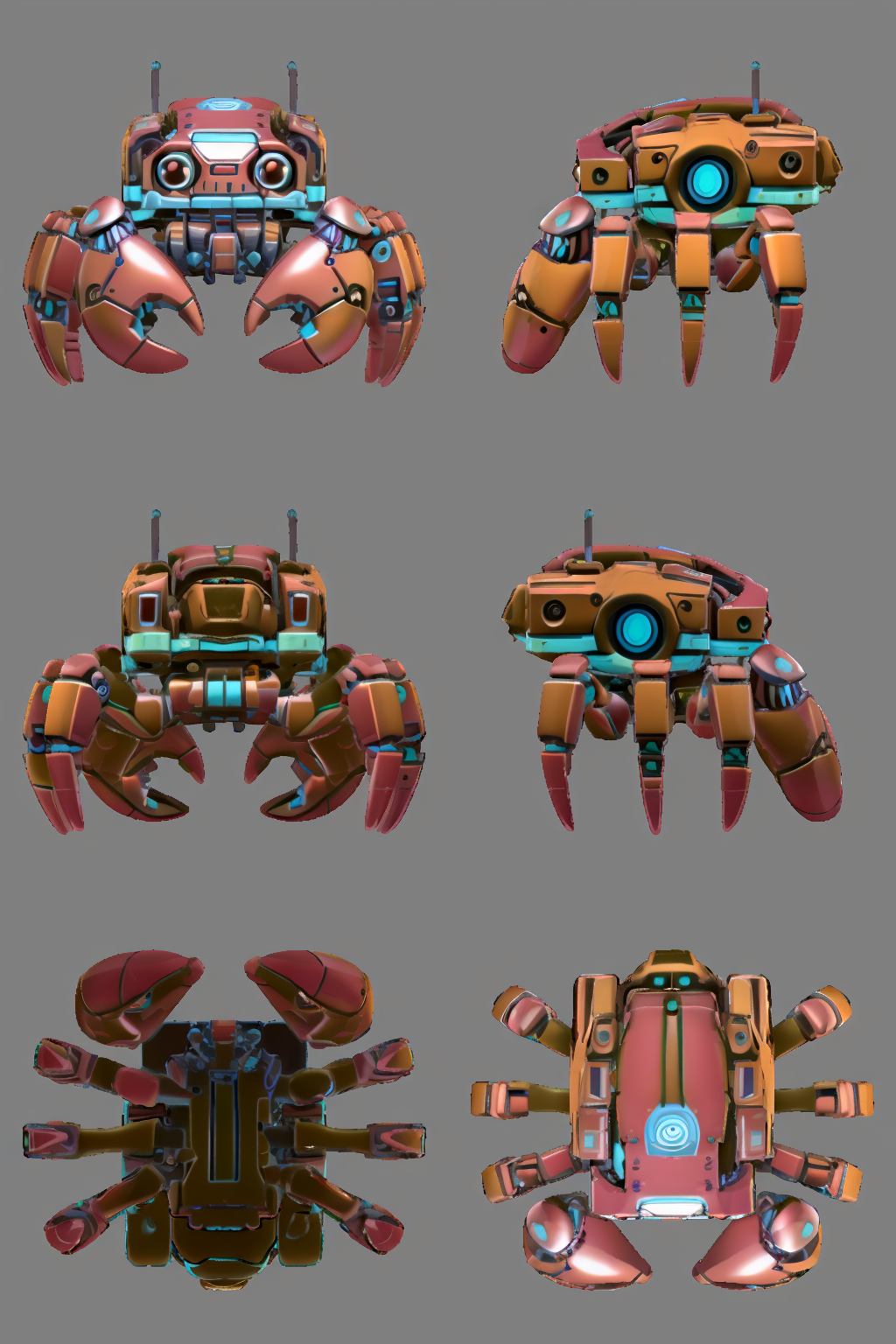}
    \includegraphics[width=0.24\linewidth]{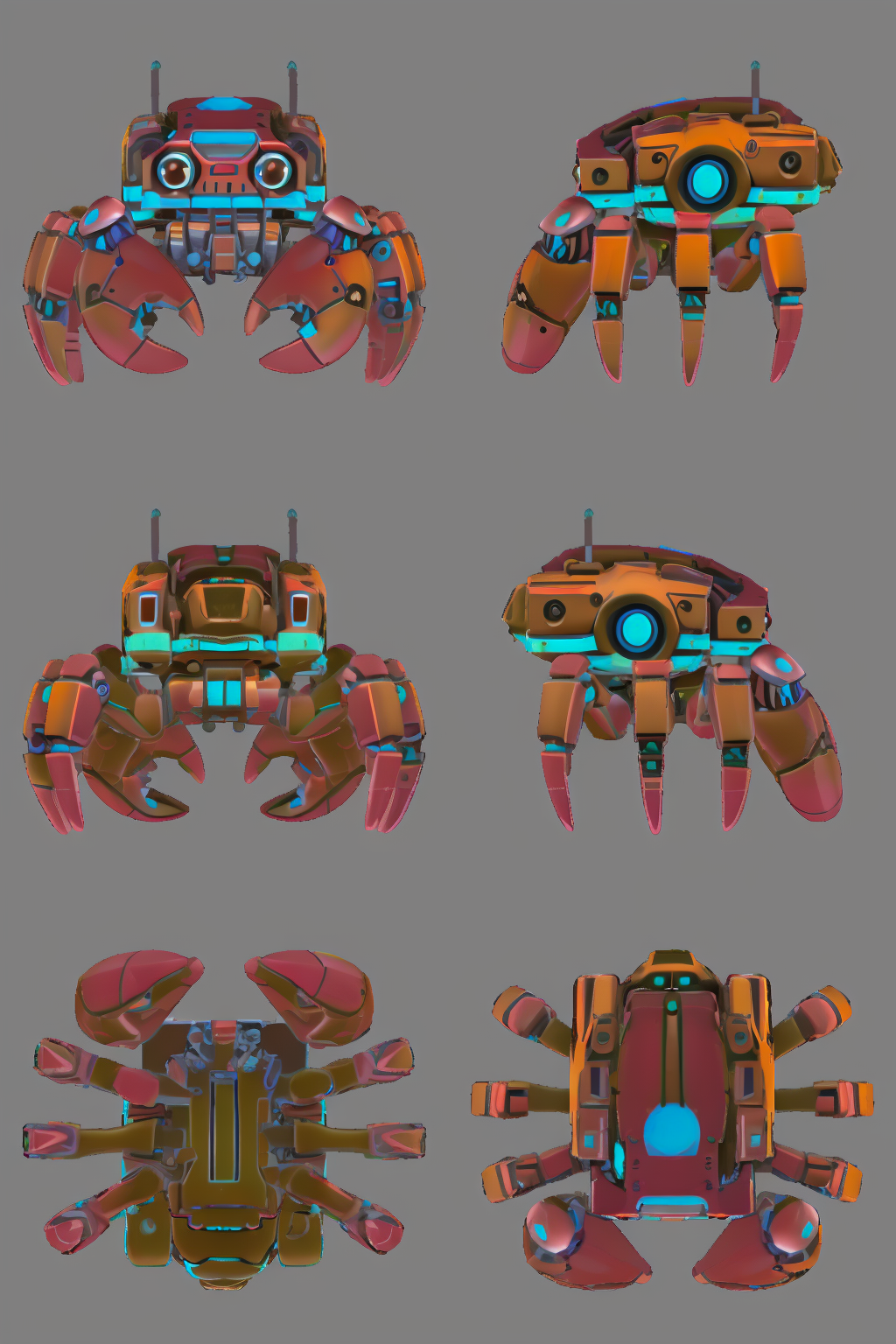}
    \includegraphics[width=0.24\linewidth]{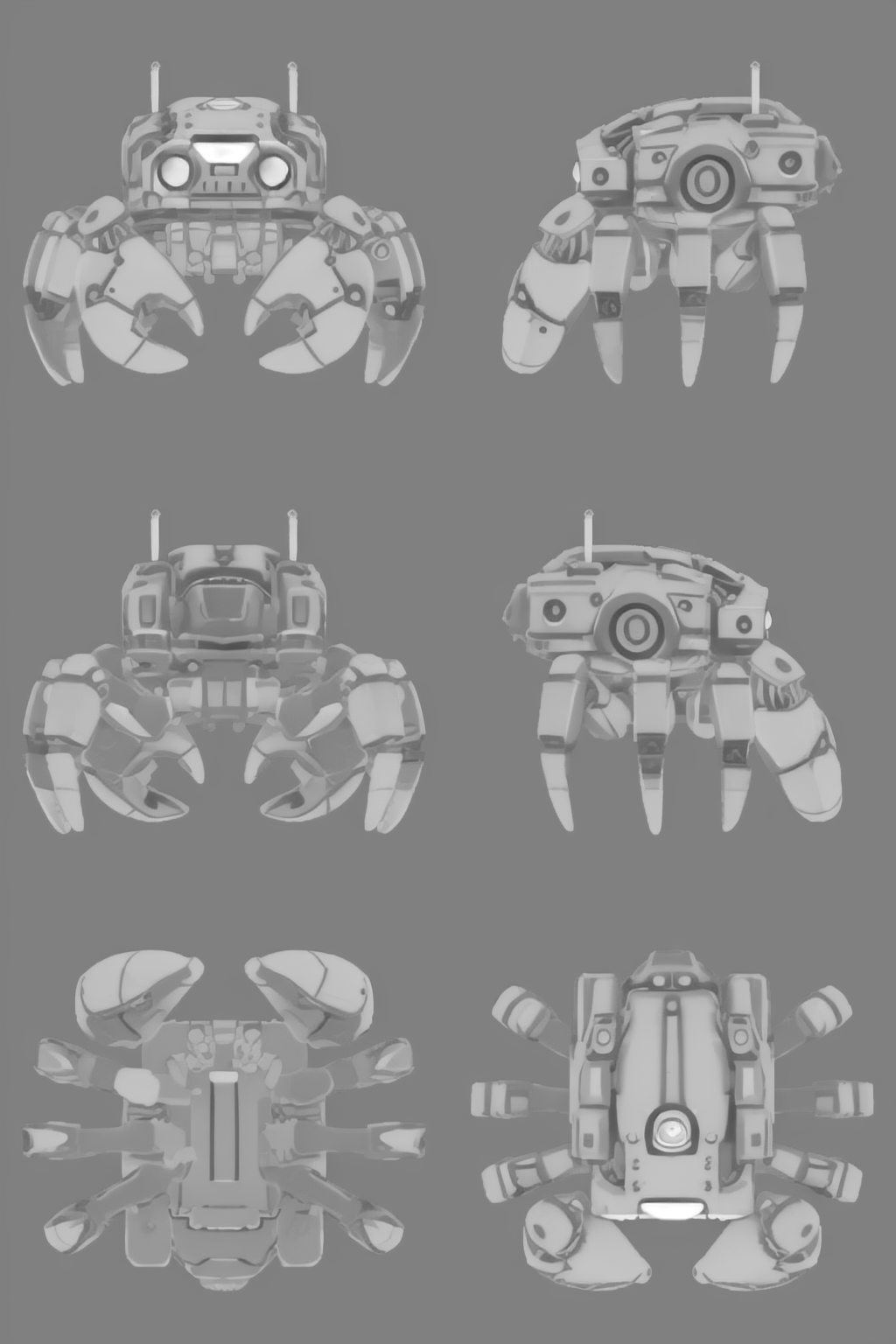}
    \includegraphics[width=0.24\linewidth]{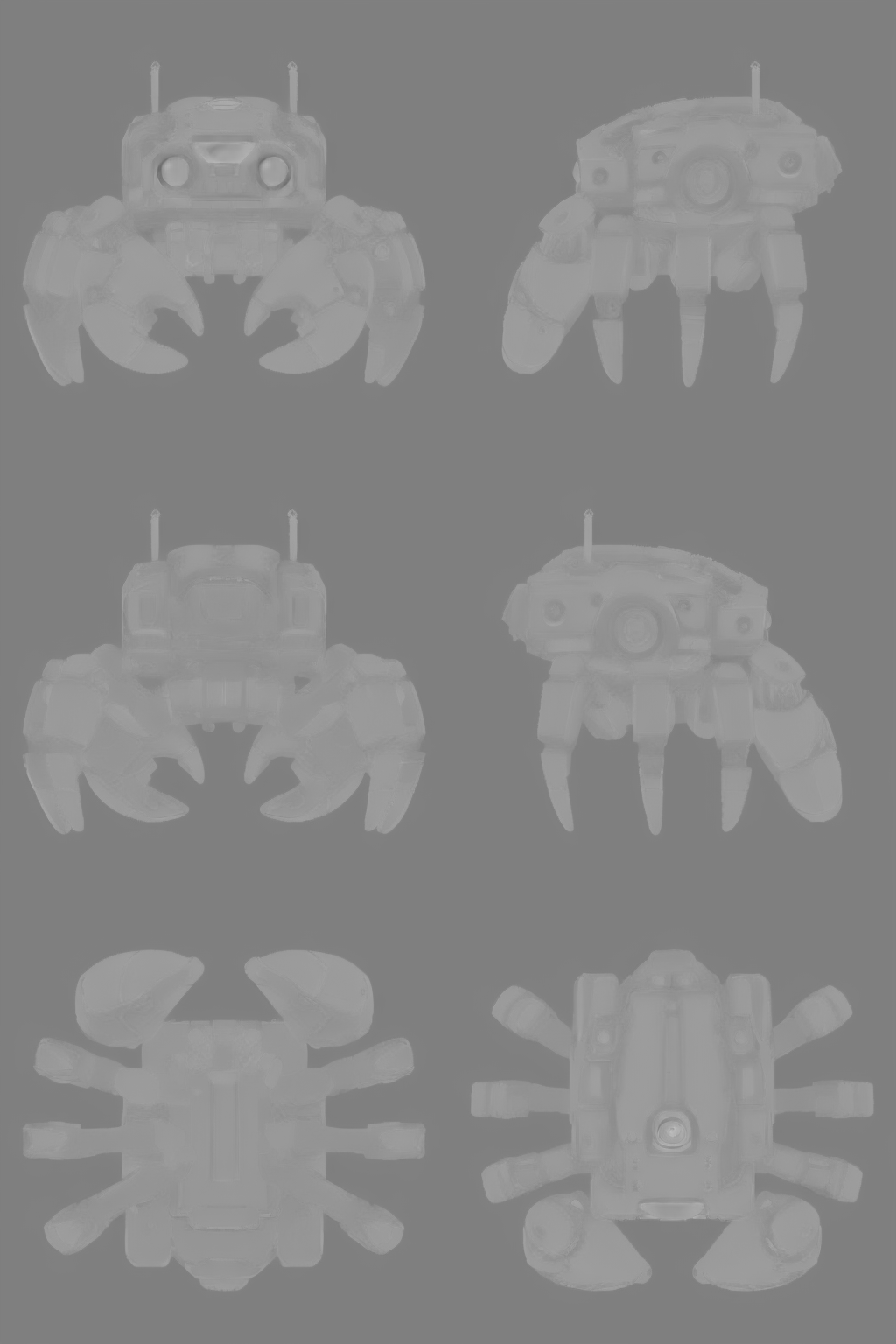}
    \caption[Multi-view Images]
    {Multi-view RGB, albedo, metallic and roughness images. }
    \label{fig:multiviewpbr}
\end{figure*}

After obtaining multi-view rgb image conditioned on multi-view position maps, we generate the corresponding multi-view PBR (Physically Based Rendering) image via a image-2-image diffusion model~\cite{rombach2022high}. Specially, taking multi-view rgb image as input, we train image-2-image diffusion models to generate corresponding multi-view PBR image. This stage includes generating multi-view albedo, metallic, and roughness maps (each sub image with resolution of 512$\times$512). Please kindly note that we train two models in multi-view PBR image generation process, where one model estimate multi-view albedo image, and one model estimate multi-view metallic and roughness maps. Examples of the generated albedo, metallic and roughness images are shown in Fig. \ref{fig:multiviewpbr}.

\subsection{High-Resolution Refinement} % jiayu
\begin{figure*}
    \centering 
    \includegraphics[width=0.65\linewidth]{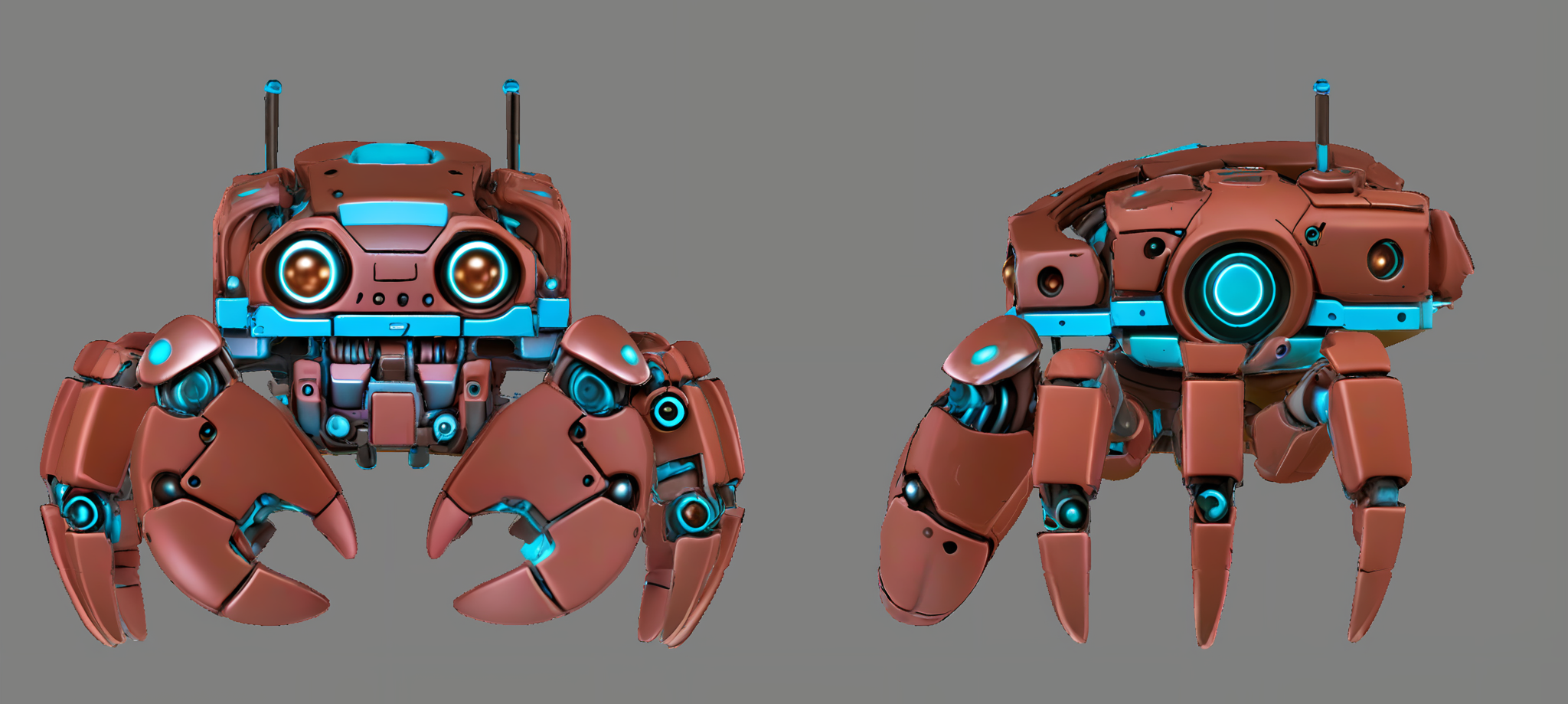}
    \caption[High-resolution albedo images]
    {High-resolution albedo images.}
    \label{fig:refined_albedo}
\end{figure*}

To further enhance visual quality, we upscale the albedo multi-view images by several iterative upscaling steps. 
Firstly, we apply two steps of image repainting, utilizing a pre-trained SDXL~\cite{podell2023sdxl} model conditioned on albedo, depth, XYZ coordinate maps, and the frontal image to upscale the albedo multi-view images to resolutions of $768 \times 768$ and $1024 \times 1024$, introducing finer details. We then use Real-ESRGAN~\cite{wang2021real} to further enhances the multi-view textures to generate detailed high-resolution albedo maps ($3072 \times 3072$), see Fig. ~\ref{fig:refined_albedo}.

\subsection{Pixel-Wise Consistency Enforcement} % jiayu, zacheng
The multi-view generation stages may produce inconsistencies at the pixel level across different views. To address this, we implement a consistent scheduler similar to TexPainter~\cite{zhang2024texpainter}, which enforces pixel-wise consistency. Specifically, the multi-view textures are firstly baked onto the 3D mesh, and a Poisson system is solved to produce seamless textures. Then, multi-view images are re-rendered and resampled into each view, ensuring consistency across different views and lighting conditions. The resampled views are used as the updated $z_0$, which is plugged into the noise scheduler of the diffusion model similar to~\cite{zhang2024texpainter}. % jiayu

\section{3D Model Data Processing} % senbo
\begin{figure*}
    \centering 

    \includegraphics[width=0.9\linewidth]{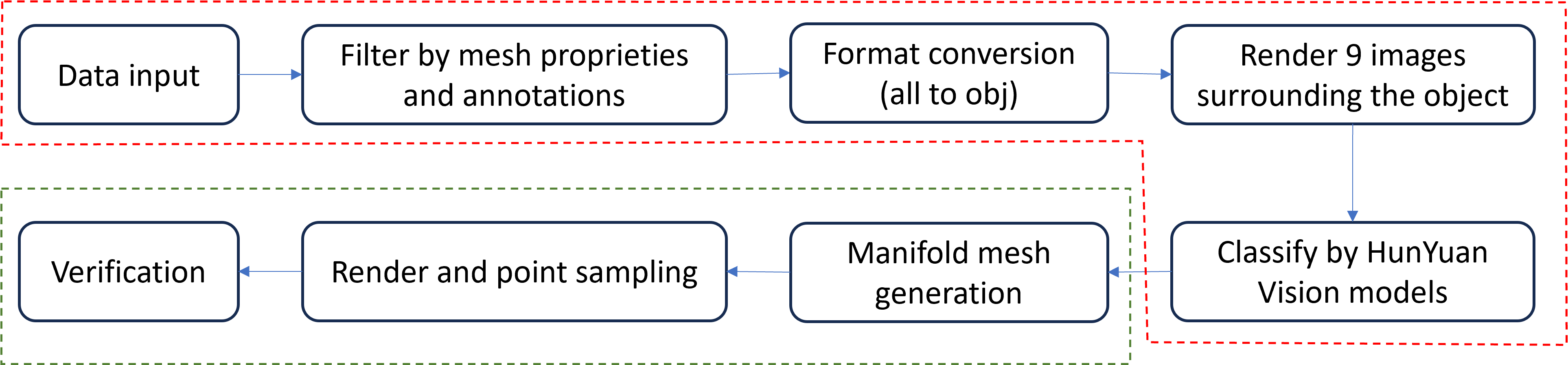}
    \caption[Data processing pipline.]
    {Data processing pipline. The procedures marked in red are one-off implementations, while the green-boxed elements demand tailored development according to the algorithmic modules deployed on the dataset, we thereby provide exampled steps for jobs described in Section~\ref{sec:3dshape} and Section~\ref{sec:texture:multiview}.}
    \label{fig:dataset:1}
\end{figure*}

\label{sec:datasets}
We collect assets (mostly 3D models) from multiple sources and preprocess them to be training compatible, including converting mesh geometry to discrete sampling of implicit functions, multi-view image generation, and PBR rendering.
Our data processing pipeline is demonstrated in Fig.~\ref{fig:dataset:1}.
We will detail each step in the processing pipeline in the following contents of this section.

\subsection{Data Origins}
\label{sec:datasets:origin}

The main data sources are:
\begin{itemize}
    \item Objaverse~\cite{deitke2023objaverse}, which is a large open-source 3D object dataset with more than $700k$ objects, we use roughly $600k$ of it;
    \item OXL~\ie Objaverse-xl~\cite{deitke2024objaverse}, which is an extension of the previous dataset Objaverse~\cite{deitke2023objaverse}, we use roughly $200k$ of it;
    \item ABO~\cite{collins2022abo}, we only use the $8k$ 3D objects in this dataset;
    \item BuildingNet~\cite{selvaraju2021buildingnet}, which contains $2k$ building models;
    \item HSSD~\cite{khanna2023hssd}, which contains roughly $13k$ object models;
    \item Toy4k~\cite{stojanov2021using}, which contains roughly $4k$ object models.
    \item Some models downloaded from the Internet, for instance, polygone dataset\footnote{\url{https://polygone.art/}}.
    \item $10k$ private data provided by our partners.
\end{itemize}

\subsection{Filter Mesh}
\label{sec:datasets:preprocess:filter}
\begin{figure}
    \centering 

    \includegraphics[width=0.9\linewidth]{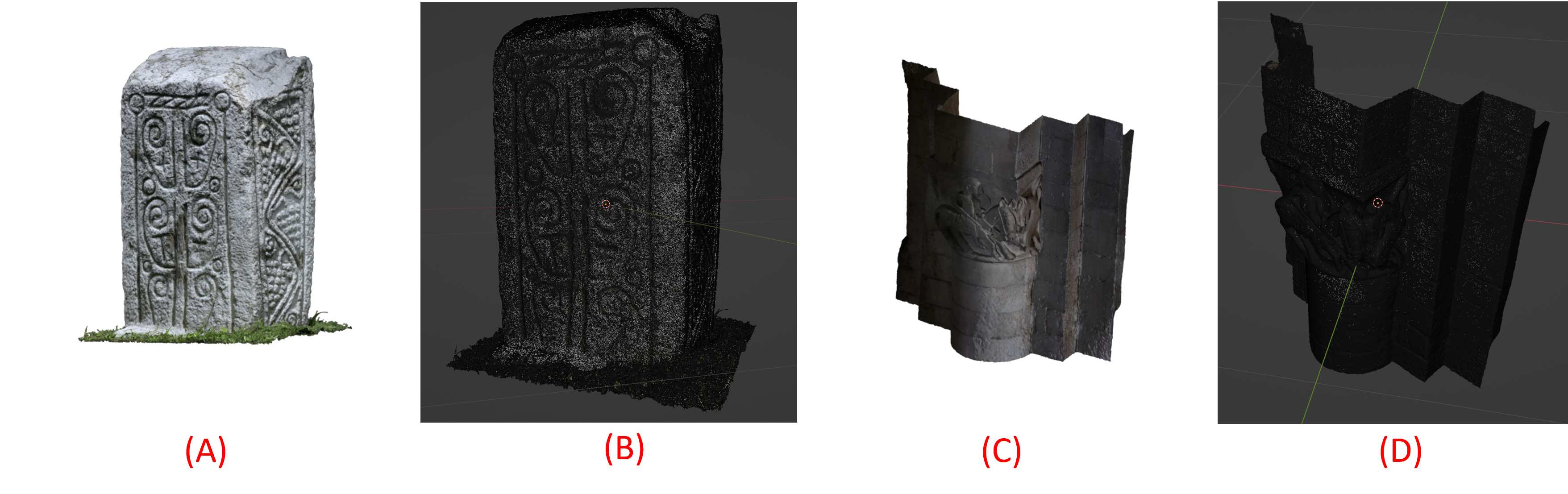}
    \caption[Scanned objects.]
    {Typical images of scanned objects in Objaverse dataset~\cite{deitke2023objaverse}. 
    (A) and (C) are rendered images of the two objects; 
    (B) and (D) are corresponding object demonstration in blender.}
    \label{fig:dataset:2}
\end{figure}

We first filter mesh using the following mesh proprieties, which are modified from the standards used in MeshXL~\cite{chen2024meshxl}:
\begin{itemize}
    \item Mesh face number, we only use meshes which face number is between $500$ and $80k$;
    \item Material number, we ignore all meshes with more than $100$ materials; material number is defined by total material number in blender\footnote{\url{https://www.blender.org/}}.
    \item Annotations of the dataset. We use annotations of Objaverse to remove scanned objects. Exampled images of scanned objects can be found in Fig.~\ref{fig:dataset:2}.
    Scanned objects are harmful to the overall 3D generative model training process in various ways:
    \begin{itemize}
        \item From (B) and (D) of Fig.~\ref{fig:dataset:2} we notice that most scanned objects possess a large number of fragmented faces;
        meshes containing such a large number of faces can only use the automatic method of unwrapping the mesh, which requires extremely large texture images to maintain reasonable mesh appearance. Hence, it's very slow in rendering.
        \item Some scanned objects do not possess an actual ``body'', like (C) in Fig.~\ref{fig:dataset:2}. 
        It's only a thin layer and not suitable for training multiple view diffusion methods like what we discuss in Section~\ref{sec:texture}.
    \end{itemize}
    \item Objects cannot be of pure color~\ie pure red or pure blue. This can be checked by checking object's material graph in Blender.
\end{itemize}

\subsection{Format Conversion}
\label{sec:datasets:preprocess:conversion}

For convenience of usage, we convert all 3D mesh formats to OBJ\footnote{\url{https://en.wikipedia.org/wiki/Wavefront_.obj_file}}. 
This is because most processing pipelines that are not part of DCC software~\ie Digital Content Creation software, cannot support read full information of complex 3D formats like GLB\footnote{\url{https://www.khronos.org/gltf/}} and FBX\footnote{\url{https://www.autodesk.com/products/fbx/overview}}. 
If we wish to scale-up for future learning-based algorithms that need to directly read information from meshes, 
we have to convert 3D format to OBJ. 
However, directly converting 3D models to OBJ format often fails, 
mostly because default format conversion function in Blender cannot correctly deal with file path of texture images. 
We thus need some extra care of some certain file types.
\begin{itemize}
    \item \textbf{MAX to OBJ} cannot be done directly, as MAX file is only supported by 3DSMax\footnote{\url{https://www.autodesk.com/products/3ds-max/overview}} and OBJ exporting function in that software cannot correctly handle objects with multiple complex materials. 
    This is because we use an extension of OBJ that supports PBR formats developed by Carla~\cite{dosovitskiy2017carla}, which is not properly supported in 3DSMax. 
    We thus first convert MAX files to FBX formats using 3DSMax, and convert FBX to OBJ using blender. 
    \item \textbf{GLB to OBJ} can be done in blender, but to get correct textures extra care is needed in rebuilding material graph structure. 
    This is because GLB specification has embedded texture images within mesh files.
    We first convert all GLB files to GLTF formats which extract texture files to disk; 
    then we go through the material graph of GLTF format and rebuild connections in new mesh.  
    \item \textbf{PMX to OBJ} can be done in blender using codes derived from Cats plugin\footnote{\url{https://github.com/absolute-quantum/cats-blender-plugin}}. PMX\footnote{\url{https://gist.github.com/felixjones/f8a06bd48f9da9a4539f}} is often used by some creators of the anime industry.
\end{itemize}

\begin{figure}
    \centering 

    \includegraphics[width=0.9\linewidth]{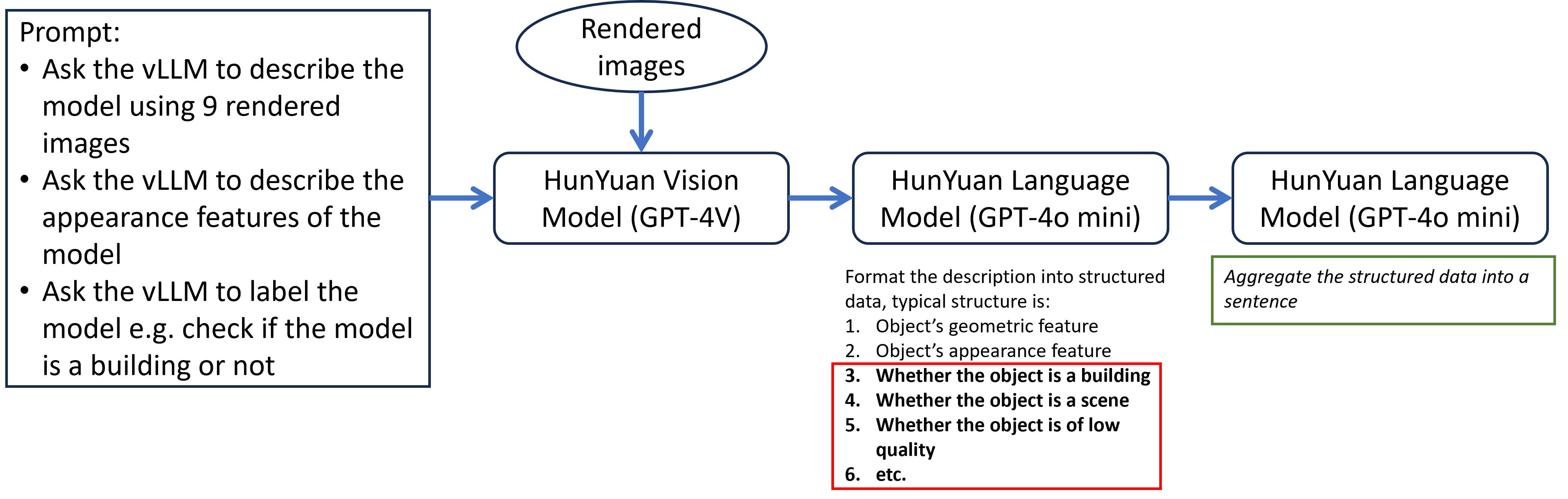}
    \caption[Mesh classification.]
    {Classify mesh using vLLM by making the vLLM model to describe the model using $9$ rendered images. Some parts of the structured data (marked with red box) can be used to classify mesh; the aggregated full sentence can be used as the caption of the mesh.}
    \label{fig:dataset:3}
\end{figure}

\subsection{Classify Mesh}
\label{sec:datasets:preprocess:classify}

The propose of this step is to eliminate low-quality mesh as thoroughly as possible, and to mark object of distinctive types. 
This provides the following advantages:
\begin{itemize}
    \item Low-quality mesh can disturb the overall training process. 
    \item Labeling mesh with its class allows us to fine-tune diffusion model on data from certain domains.
\end{itemize}
The filtering process descirbed in this section is modified from the process used in MeshXL~\cite{chen2024meshxl}, as shown in Fig.~\ref{fig:dataset:3}. 
After this process, we render $9$ images surrounding the mesh, 
and use the HunYuan vision model\footnote{\url{https://cloud.tencent.com/document/product/1729/104753}} to filter the mesh by these rendered images. 
Note that you can substitute this vision model with any other vLLM models, such as GPT-4V\footnote{\url{https://openai.com/index/gpt-4v-system-card/}}.
As shown in Fig.~\ref{fig:dataset:3}, the text prompt guides the vLLM to describe the appearance of the object, and check if the object is of certain classes. 
Then, we use LLM~\cite{sun2024hunyuan} to convert unstructured text into a structured format with structures similar to JSON.
Labels in these structured texts can be used as the object's class.

\subsection{Generate watertight mesh}
\label{sec:datasets:preprocess:filter}

Generating watertight mesh, which is essential for 3D-DiT training, is generating an envelope tightly conforming to the model's exterior geometry, 
rather than using both interior and exterior geometry. 
The latter, like Manifold~\cite{huang2018robust,huang2020manifoldplus} and other works~\cite{portaneri2022alpha} that can provide similar results, is not suitable for 3D-DiT training.

We use similar watertight mesh generation method from 3DShape2VecSet ~\cite{zhang20233dshape2vecset}, 
which is adopted from Stutz's work~\cite{stutz2020learning} and used in occupancy networks~\cite{mescheder2019occupancy}\footnote{We directly use implementation from \url{https://github.com/autonomousvision/occupancy_networks/}.}.
The method is based on TSDF fusion of a group of depth map rendering around the object.
However, this method will slightly vary the exterior shape of the mesh due to the following reasons:
\begin{itemize}
    \item The method calculates closing of each depth map, which fills small gaps on depth image but remove some details. Distortion caused by such reason can be reduced by increasing the definition of depth image, as closing of the image is often calculated using fixed window size.
    \item The method applies a bias that is half voxel size. Distortion caused by such reason can be reduced by high-definition SDF volume. 
\end{itemize}
However, increasing the definition of the SDF volume will lead to a mesh with particularly large face number, which requires the pipeline to provide a decimated mesh. 
It's worth noticing that most decimating methods in DCC software uses QEM triangle decimation~\cite{garland1997surface}, which will sometimes provide ill-formed faces \ie the shape of the face may not be suitable for further developments.
We therefore also recommend using ACVD~\cite{audette2011approach,valette2008generic,valette2004approximated} in the decimating step.
We have also developed a baking tool based on baking function in Blender to provide texture for watertight mesh.

\subsection{Rendering and Sampling}

All meshes are normalized to a tightly coupled (radius is 1) bounding sphere using Welzl's algorithm~\cite{welzl2005smallest}. We prefer not to use bounding boxes because arbitrarily rotating objects within them may cause the objects to extend beyond the confines of the box. 
Rendering is done in Blender using EEVEE\footnote{\url{https://docs.blender.org/manual/en/latest/render/eevee/introduction.html}} renderer, while sampling is done using trimesh\footnote{\url{https://trimesh.org/}}. 
We sample three groups of points, each with $500k$ points, which is similar to sampling approach used by For 3D geometry compression model, we build upon CraftsMan~\cite{li2024craftsman}, who adopts structures introduced in 3DShape2VecSet ~\cite{zhang20233dshape2vecset}:
\begin{itemize}
    \item We perform uniformly random sampling within the circumscribed cube of the bounding sphere, yielding SPACE points.
    \item We sample a group of points just on surface of the watertight mesh, yielding SURFACE points. We compute Gaussian curvature of each vertex on the mesh and use this curvature as importance of each area on the mesh: more points will be sampled on areas with higher curvature. 
    \item We perform uniformly random sampling on the surface of the mesh, and add a small bias to coordinates of each sampled point, yielding NEAR-SURFACE points.
\end{itemize}

\section{Experiments}

% \subsection{Quantitative Results}

% \subsection{Qualitative Results}

\begin{figure*}
    \centering 
    \includegraphics[width=\linewidth]{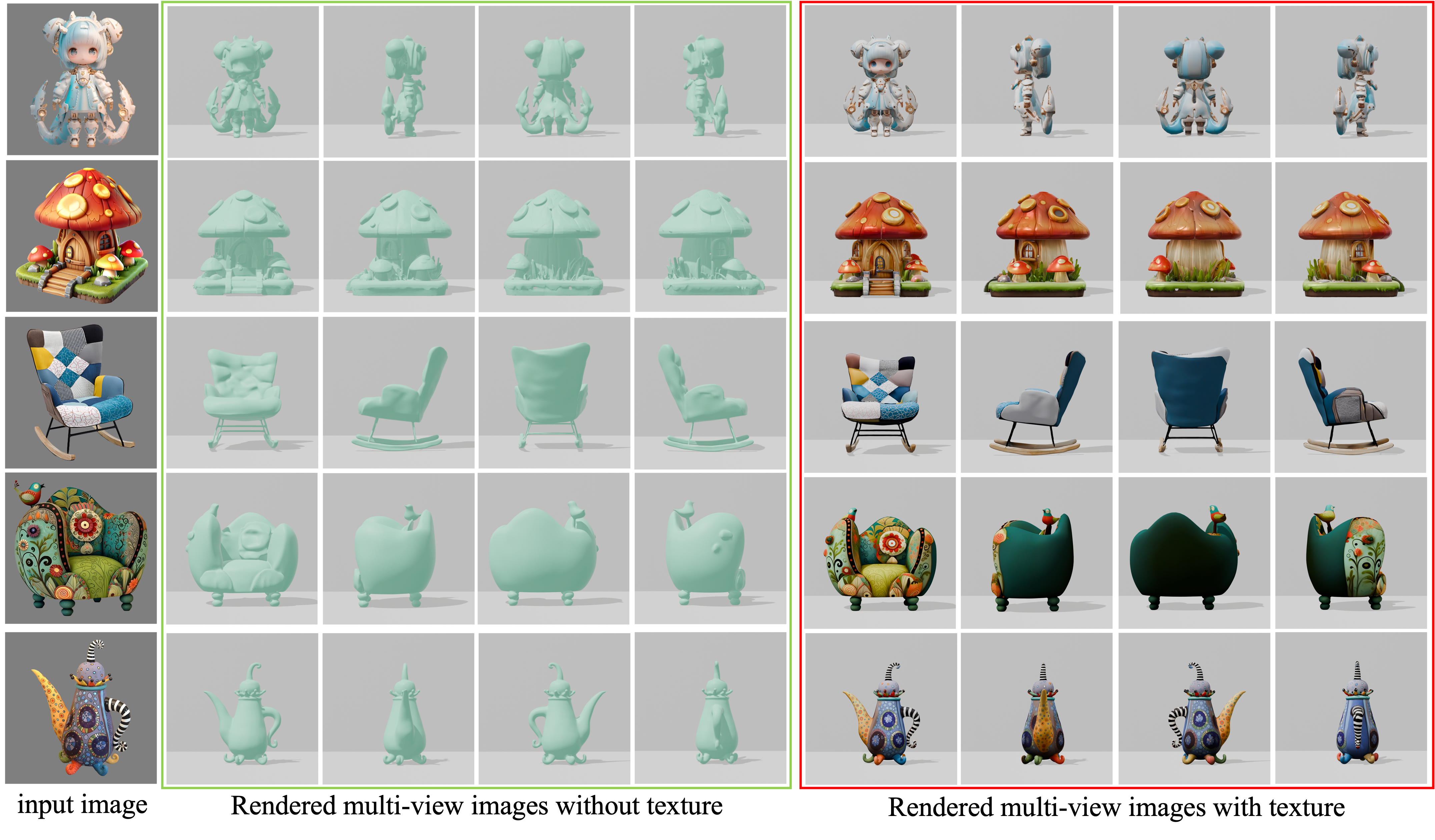}
    \caption[3D generation with image as input]
    {Visual results with color image as input, the green areas show the rendered multi-view images without textures and the red areas show the rendered multi-view images with textures.}
    \label{fig:image-2-mesh-res}
\end{figure*}

\begin{figure*}
    \centering 
    \includegraphics[width=\linewidth]{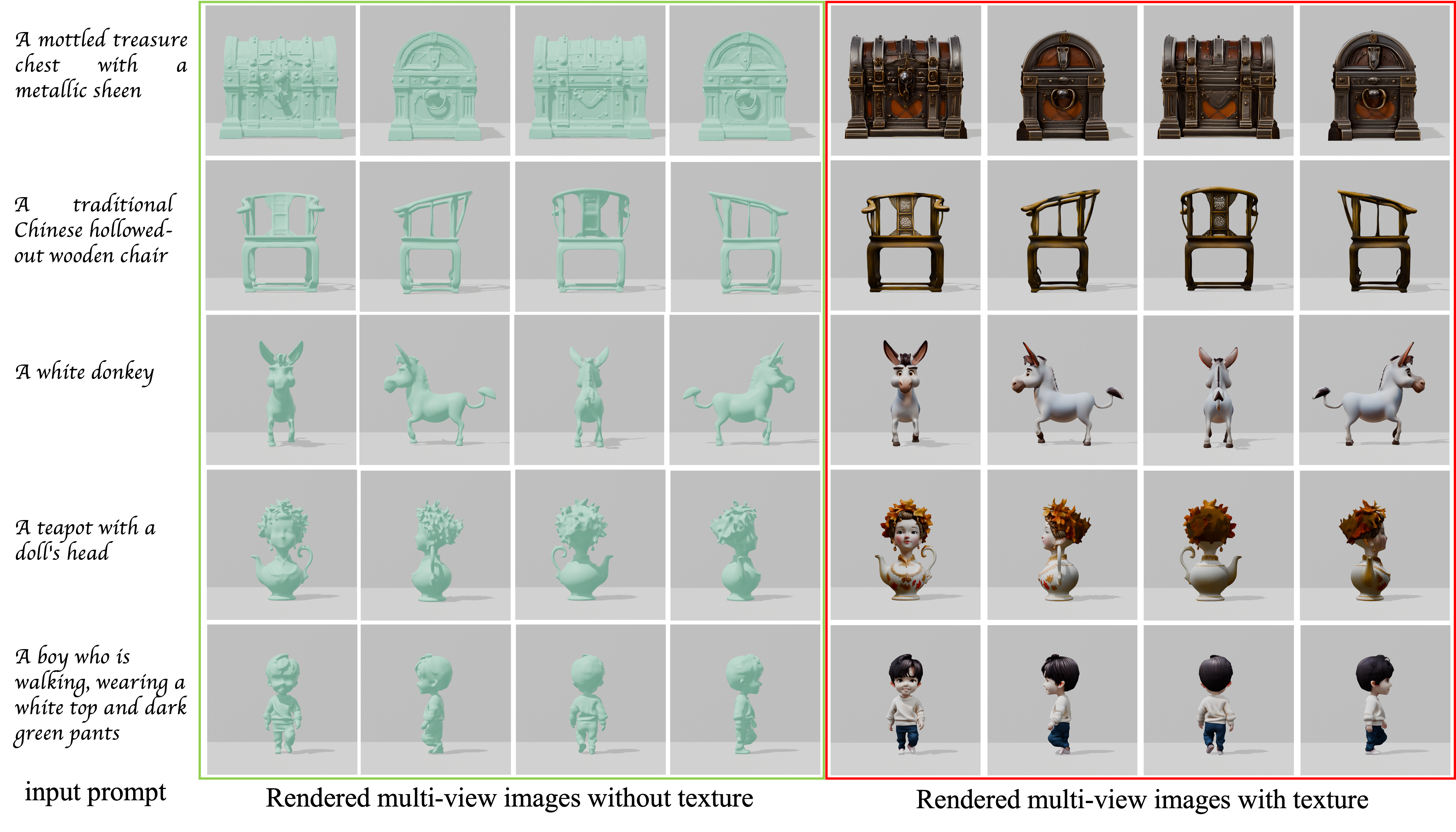}
    \caption[3D generation with prompt as input]
    {Visual results with prompt as input, the green areas show the rendered multi-view images without textures and the red areas show the rendered multi-view images with textures.}
    \label{fig:text-2-mesh-res}
\end{figure*}

Fig.~\ref{fig:image-2-mesh-res} and Fig.~\ref{fig:text-2-mesh-res} illustrate the results of 3D generation with the prompt and the image as input. As shown in these figures, which shows that our Pandora3D system could faithfully recover the shape and texture with fine-grained details and produce neat space without any floaters.

\section{Conclusion}

In this report, we present Pandora3D, a framework designed for high-quality 3D shape and texture generation. 3D shape generation and texture generation are proposed in Pandora3D. In specific, the 3D shape generation utilizes a Variational Autoencoder (VAE) to encode implicit 3D geometries into a latent space; then, a diffusion network is used to generate latents conditioned on input prompts, with modifications aimed at enhancing the model's capacity. Meanwhile, we also explore an alternative Artist-Created Mesh (AM) generation approach, which shows promising results for simpler geometries. The texture generation process is multi-staged, starting with the generation of frontal images, followed by multi-view images generation, RGB-to-PBR texture conversion, and high-resolution multi-view texture refinement. A novel consistency scheduler is integrated into every stage of this process to ensure pixel-wise consistency among multi-view textures during inference, leading to seamless integration. Experiment results demonstrate the effectiveness of Pandora3D handling of diverse input formats to produce high-quality 3D content. 
{\small
\bibliographystyle{ieee_fullname}
\bibliography{references}
}

\end{document}